\documentclass[
 reprint,
superscriptaddress,
 amsmath,amssymb,
 aps,
pre,
floatfix,
 lengthcheck
]{revtex4-1}
\usepackage[utf8]{inputenc}
\usepackage[T1]{fontenc}
\usepackage[version=4]{mhchem}
\usepackage{graphicx}
\usepackage{placeins}
\usepackage[usenames, dvipsnames]{xcolor}
\usepackage{url}

\begin{document}

\title{Machine learning in interpretation of electronic core-level spectra}

\author{J.~Niskanen}
\email{johannes.niskanen@utu.fi}
\affiliation{University of Turku, Department of Physics and Astronomy, FI-20014 Turun yliopisto, Finland}
\author{A.~Vladyka}
\affiliation{University of Turku, Department of Physics and Astronomy, FI-20014 Turun yliopisto, Finland}
\author{J.~A.~Kettunen}
\noaffiliation{}
\author{Ch.~J.~Sahle}
\email{christoph.sahle@esrf.fr}
\affiliation{European Synchrotron Radiation Source,71 Avenue des Martyrs F-38000 Grenoble, France}

\begin{abstract}
Electronic transitions involving core-level orbitals offer a localized, atomic-site and element specific peek window into statistical systems such as molecular liquids. Although formally understood, the complex relation between structure and spectrum -- and the effect of statistical averaging of highly differing spectra of individual structures -- render the analysis of an ensemble-averaged core-level spectrum complicated. We explore the applicability of machine learning for molecular structure -- core-level spectrum interpretation. We focus on the electronic Hamiltonian using the \ce{H2O} molecule in the classical-nuclei approximation as our test system. For a systematic view we studied both predicting structures from spectra and, vice versa, spectra from structures, using polynomial approaches and neural networks. We find predicting spectra easier than predicting structures, where a tighter grid (even unphysical) of the spectrum improves prediction, possibly inviting for over-interpretation of the model. The accuracy of the structure prediction worsens when moving outwards from the center of mass of the training set in the structural parameter space, which can not be overcome by model selection based on generalizability.
\end{abstract}

\maketitle
\section{Introduction}
Machine learning (ML) is becoming a standard tool in research questions where numerous repeated evaluations of a complicated model are required. In such cases using ML as an emulator may provide enormous relief in computational burden \cite{hutson2020}. In the context of physics, ML means building a fundamentally unphysical model such as a neural network (NN) to describe data and to make predictions for new input. Light evaluation cost of a model would then allow for numerous predictions to be performed to simulate a statistical average, or to iteratively solve a given problem.
\par
 Statistical studies of core-level spectra fit to the category of repeatedly evaluated physical models by their definition. Core-level spectroscopic methods can be used for characterization of materials and their function on the atomic level \cite{siegbahnESCAbook2,stohrNEXAFSbook,schulke2007book}. The spectra reflect transitions between electronic states and, therefore, their energies and transition probabilities are dictated by quantum mechanics. The benefit of using core-level excitations is, that the initial orbital for the electronic transition is localized at one atomic site in the system, which means that the process is a local probe, although the measured signal represents an explicit ensemble average. In interpretations it is a typical approximation to consider only the electronic system (fixing the positions of the nuclei), which renders the underlying quantum mechanics, and the resulting spectra, to be functions of the atomic coordinates. Although the connection of the two is clear, the interpretation of these spectra in terms of underlying atomistic structure or changes therein is not trivial. One reason for this difficulty rises from complexity, as the data linking the individual structural parameters to the line intensities, can be heavily scattered due to statistical phenomena \cite{scirep2016,pre2017}. 
\par
The idea of collaborative action of the structural parameters as predictors of X-ray spectra raises hope that more structural information from experiments could be obtained with simulations of X-ray spectra and ML, as found in the works of Refs. \cite{fraenkel-1,fraenkel-2,fraenkel-3}. In addition, application of the machine learning approaches demonstrated great performance in the prediction of the electronic structures of the atomic systems \cite{Chandrasekaran2019} as well as in the prediction of UV/Vis \cite{gosh2019} and X-ray absorption spectra \cite{Carbone2020} of simple molecules. In this work, we study the use of ML to yield predictions for oxygen K-edge spectra and structures for the \ce{H2O} molecule from a statistical simulation. The problem belongs to the supervised regression learning category as two separate tasks: (i) training ML on known data to predict spectra for new configurations, and (ii) training ML on known data to predict configurations for new spectra.

\begin{figure}
\includegraphics[width=\linewidth]{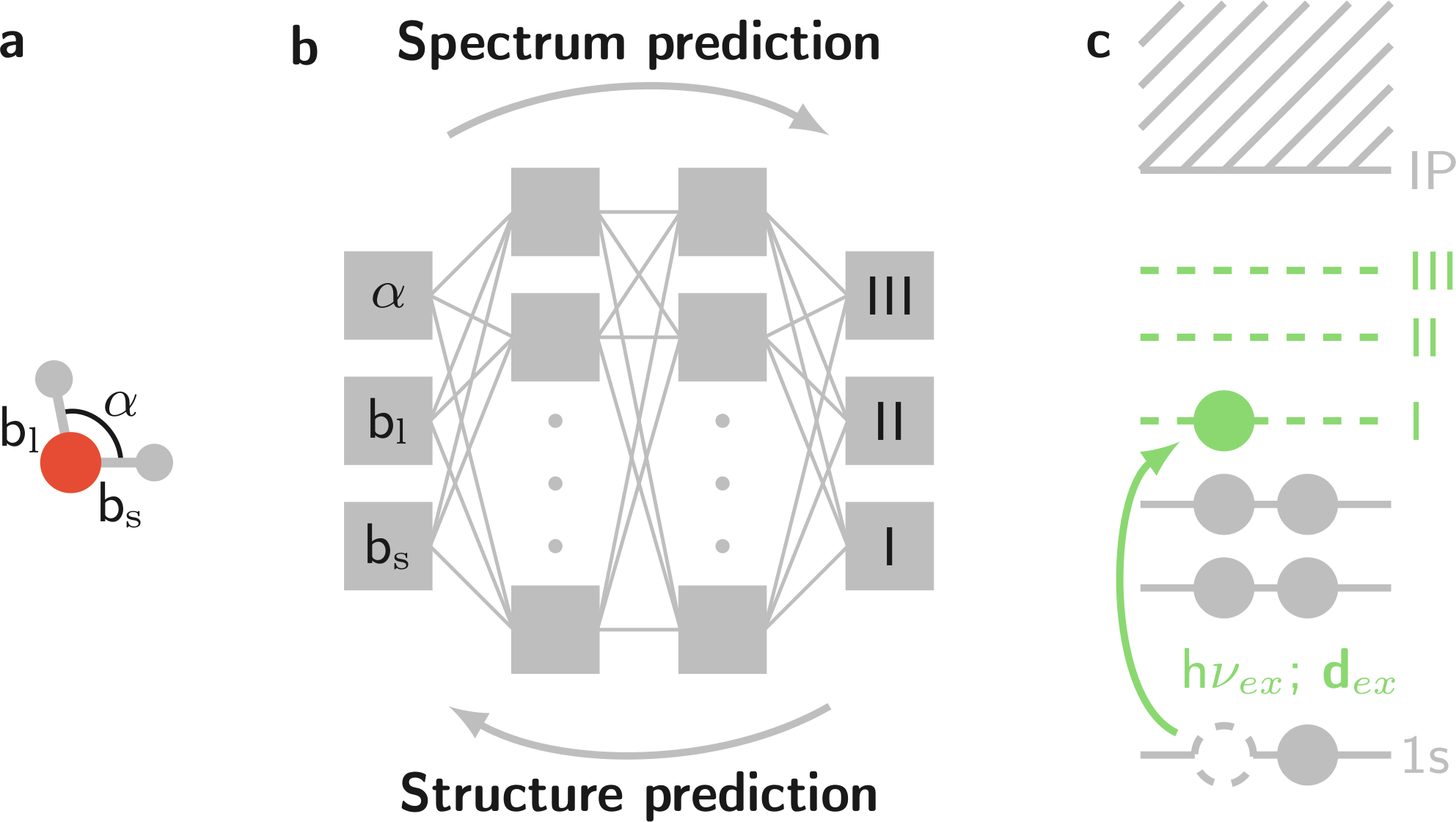}
\caption{\label{Schematic} The \ce{H2O} molecule and its structural parameters (a), linked by a multilayer feed-forward perceptron (b) (polynomial models were also studied), to the lines in the O K-edge absorption spectrum (c). Intensity in spectral regions of interest (ROI) I, II, and III is studied. A neural network for the reversed direction addresses the structure prediction from the ROI intensities.}
\end{figure}
We benchmark these ideas for a simple system, H$_2$O in the gas phase, which provides a manageable system both for calculations and for human intuition, owing to the few structural degrees of freedom, the two OH bonds and one H-O-H angle. Although not achieved here, our work is motivated by the interpretation of structures from spectra in condensed phases, where the classical-nuclei model is the contemporary standard. Therefore, we deliberately choose to use the classical framework for the nuclear subsystem even though more accurate calculation with quantum vibrations would also be possible \cite{couto2017,keith2018}. Moreover, we generate the structures for ML by sampling {\it ab initio} molecular dynamics trajectories, instead of sampling the ground state probability distributions in the harmonic approach, which would have been a suitable option in this case \cite{Leetmaa2010}.

\section{Methods}
We studied four prediction tasks based on simulated structures and spectra: structure prediction from a full (75 points) and from a coarsened spectrum, and spectrum prediction (full and coarsened) from the structure. In each case, we trained ML systems with structure--spectrum data based on {\it ab initio} molecular dynamics (AIMD) simulations and spectrum evaluation for obtained structures. We study the effect of spectral binning by presenting the data either on a 0.1-eV-spaced grid and by coarsening it to a few values by integration over regions of interest (ROI) that were chosen to match the minima of the ensemble averaged spectrum. We first trained the models with data obtained at initial kinetic energy equivalent to 480~K (low-E set). To study the generalizability of the estimators, we repeated the structural and spectral simulation at initial kinetic energy equivalent to 10000~K (high-E set). This study provided us with a data set of the same size with larger coverage of the configuration space. 
\par
We applied mean-standard-deviation normalization (based on values of the training set) for both input and output variables and applied the inverse transform to the output after prediction. Contrary to the coarsened spectral ROI data, the tight-gridded data has channels of near zero intensity and no spectral information (e.g. below 533 eV), which caused instabilities with independent normalization of channel intensities. Therefore we standardized the tight-gridded spectral data by using the mean and the standard deviation of all channels and spectra collectively, instead of individual transformation for each channel.
\par
We divided the data to training sets (80\%), and to test sets (20\%) and used cross-validation (CV) to evaluate the goodness of a particular model -- and choose the best-performing model from this hypothesis space. Then we evaluate the final learning ability of the model by using the test set. For NN with rectified linear unit (ReLU) activation, we ran a 3-dimensional grid-search model selection. We studied the regularizing parameter alpha ($\alpha=10^n$, $n=-10,-9,\ldots,4$), as well as the network depth (2,$\ldots$,5) and width (5, 10, 50, 100, 200, 500) to obtain 360 NN models in each case. We used the \texttt{scikit-learn} \cite{scikit-learn} package, and the Adam \cite{adam} solver, with upper limit for the number of iterations of $10^7$. For comparison, polynomial models up to the 9th order were used, with regularizing parameter $\alpha=10^n$, $n=-10,-9,\ldots,4$. Here, a singular-value decomposition algorithm was used.
\par
For the structures we performed independent 100 1-ps-long AIMD runs sampling the NVE ensemble (initial T=480~K, $\Delta$t=0.5~fs) to sample a set of phase space points (basis TZV2P-MOLOPT-GTH \cite{molopt}, pseudopotentials: GTH-PBE \cite{gth}, exchange-correlation potential PBE \cite{pbe}, cutoff 300~Ry). From each of these trajectories (first 250~fs ignored) 100 structures were sampled randomly. As structural parameters we use the bond angle, the length of the shorter bond, and the length of the longer bond as from this information the molecular electronic Hamiltonian, and its excitation spectrum, are uniquely defined. For the high-E data, the simulation time step $\Delta$t=0.1~fs was used.
\par
We used transition-potential density functional theory (TP-DFT) \cite{Triguero1998} in the half-core-hole approximation to evaluate 10000 O K-edge spectra (energies and intensities) for the structural data sets. We apply an explicit $\Delta$-DFT energy correction for each spectrum by finding the shift for the lowest excited state. The $\delta$-peak spectra were convoluted by a Gaussian of 1~eV full-width-at-half-maximum (FWHM). The aug-cc-pV5Z \cite{dunning,dunning2} basis was used for all atoms in the spectrum evaluation. Calculations for structures and spectra were carried out using the CP2K software \cite{cp2k}.
\section{Results}
\begin{figure}
\includegraphics[width=\linewidth]{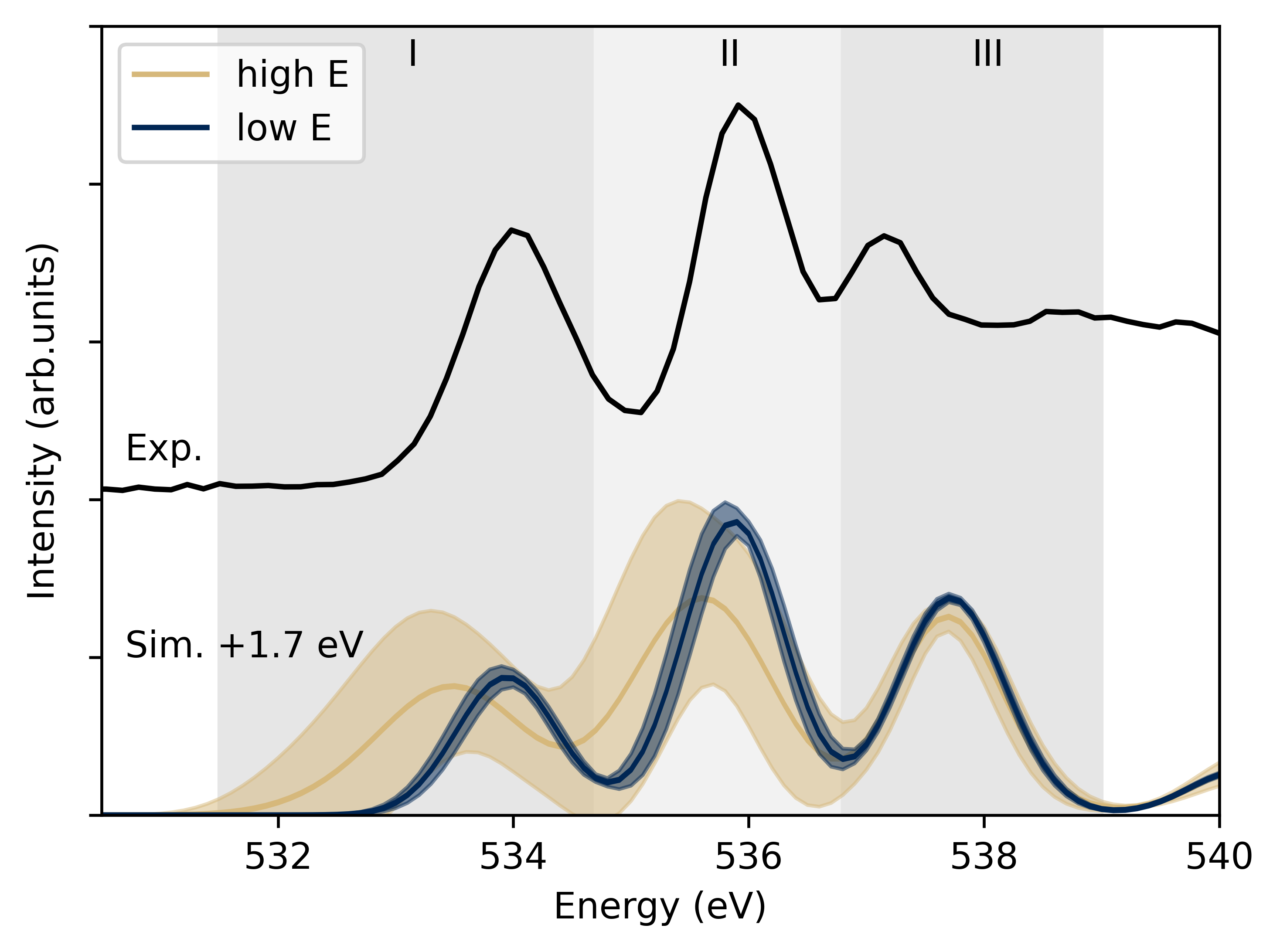}
\caption{\label{figure1} Simulated low-E and high-E excitation spectra of \ce{H2O} compared to experimental O K-edge excitation spectrum. The standard deviation ($\pm\sigma$) of the simulation results are shown as shading.}
\end{figure}

\begin{figure}
\includegraphics[width=\linewidth]{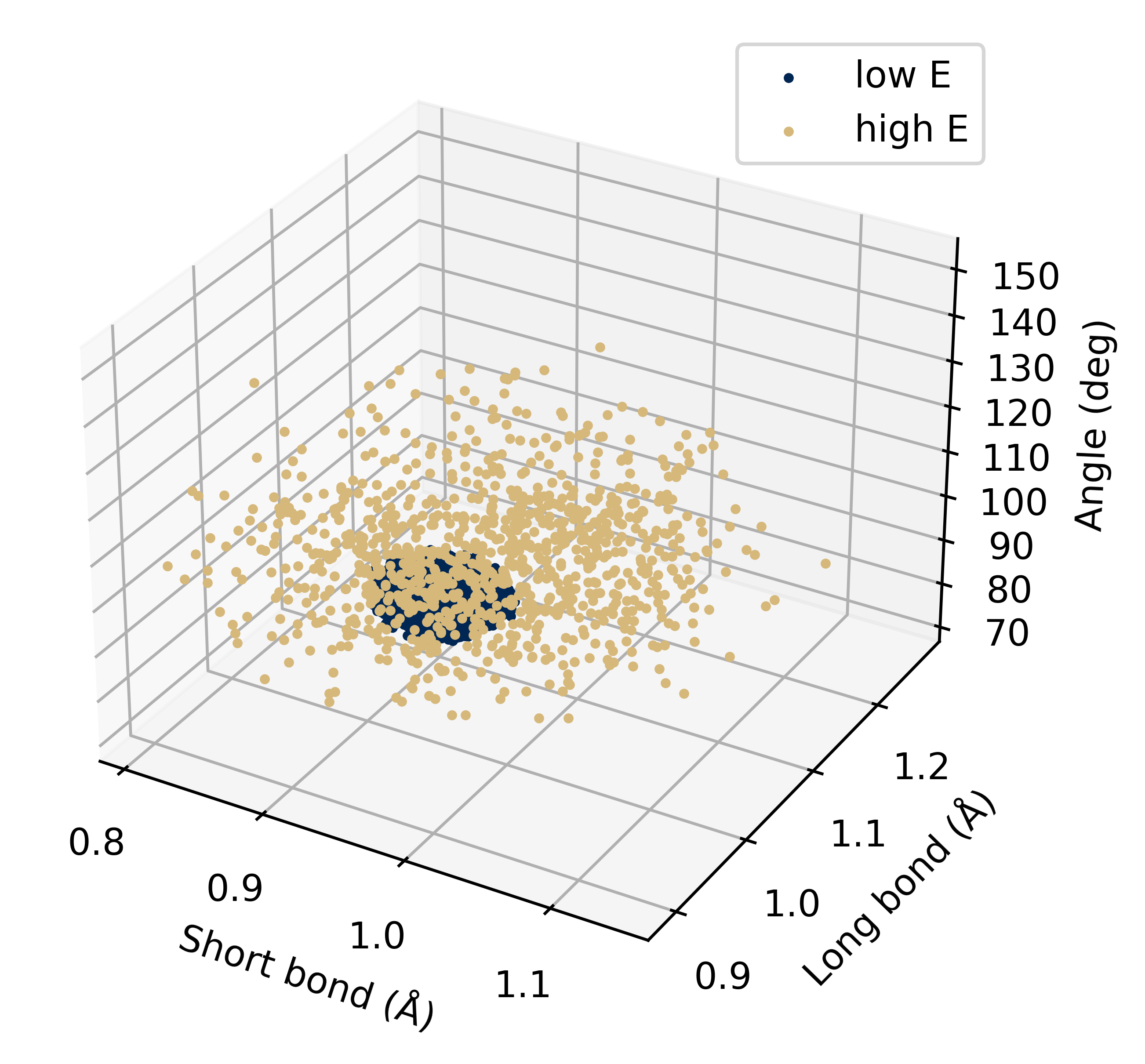}
\caption{\label{PhaseSpace} Sampling of the phase space in the two simulations. For clarity, only 1000 points are shown for the high-E simulation.}
\end{figure}

Figure \ref{figure1} shows the average of simulated K-edge spectra of \ce{H2O} for the two data sets, in comparison experimental gas phase spectrum (black solid line; raw data from \cite{hitchcockwater,hitchcockdatabase}). Figure \ref{PhaseSpace} shows the corresponding structural data in the three-dimensional phase space. Higher energy allows the system to cover a wider region in the configuration space. The results are divided to two sections according to the spectral binning (ROI or tight grid), and the optimal models in each studied case are listed in Table \ref{tab:best_comparison}. For our set of hyperparameters, we observed a variety of NN architectures to converge with similar goodness. Moreover, for different grid search runs, the `best' configuration was observed to vary with score differences within the tolerance interval, which we interpret to rise from the stochastic nature of the used algorithms. Learning curves for the best-performing models are presented in the Appendix, and show convergence with training set size in all studied cases.
\par
\par
\begin{figure}
\includegraphics[width=\linewidth]{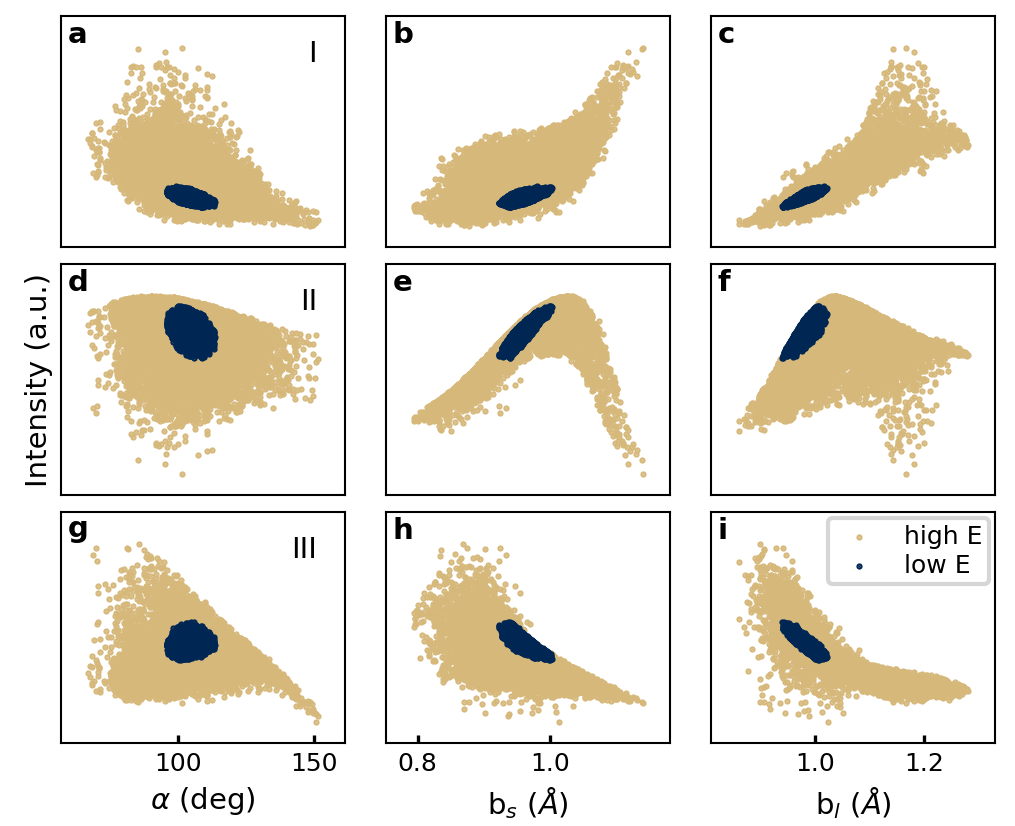}
\caption{\label{figure2}The ROI intensities in regions (I, II, III) plotted against the underlying structural parameters for the simulation of O K-edge spectrum of \ce{H2O} of these structures.}
\end{figure}
\par
\begin{table*}[ht]
\caption{\label{tab:best_comparison}Typical best-performing models found in the work; several NNs performed equally well and the result depended on the stochastic search. The average cross-validation mean squared error (MSE) for the normalized prediction output values is also provided. For structure prediction, the optimal parameter-limited models are given. For details, see text.}
\begin{tabular}{l l | l l | l l}
& & 3 ROIs & & 75-point grid & \\
Prediction & Training Data & Best Model & CV-MSE & Best Model  & CV-MSE \\
\hline
Spectra         & low-E  & 3rd order Poly, $\alpha=10^{-1}$ & 0.027 & 4th order Poly, $\alpha=1$ & 0.0003\\
Spectra         & high-E & 9th order Poly, $\alpha=10^{-1}$ & 0.0016 & 9th order Poly, $\alpha=10^{-1}$ & 0.0004\\
Structure       & low-E  & 2 $\times$ 500, ReLU, $\alpha=10^{-4}$ & 0.084 & 3 $\times$ 200, ReLU, $\alpha=10^{-1}$ & 0.0044\\
Structure       & high-E & 4 $\times$ 200, ReLU, $\alpha=10^{-4}$ & 0.053 & 2 $\times$ 500, ReLU, 
$\alpha=10^{-2}$& 0.001\\
Structure (lim) & low-E  & 3 $\times$ 100, ReLU, $\alpha=10^{-1}$ & 0.083 & 2 $\times$ 50, ReLU, $\alpha=10^{-1}$ & 0.0053\\ 
Structure (lim) & high-E & 3 $\times$ 100, ReLU, $\alpha=10^{-7}$ & 0.055 &  3 $\times$ 100, ReLU, $\alpha=10^{-1}$ & 0.0013\\
\hline
\end{tabular}
\end{table*}

\subsection{Coarsened spectral data}
The studied ROIs are indicated by gray shading in Figure \ref{figure1}. Figure \ref{figure2} shows the ROI intensities as a function of the three structural parameters: the data is transferred into structural-parameter --ROI-intensity representation as in Ref.~\citenum{pre2017}. This view, however, misses the collaborative effect of parameters as the analysis focuses on individual structure--ROI dependencies. 
\par
\begin{figure}
\includegraphics[width=\linewidth]{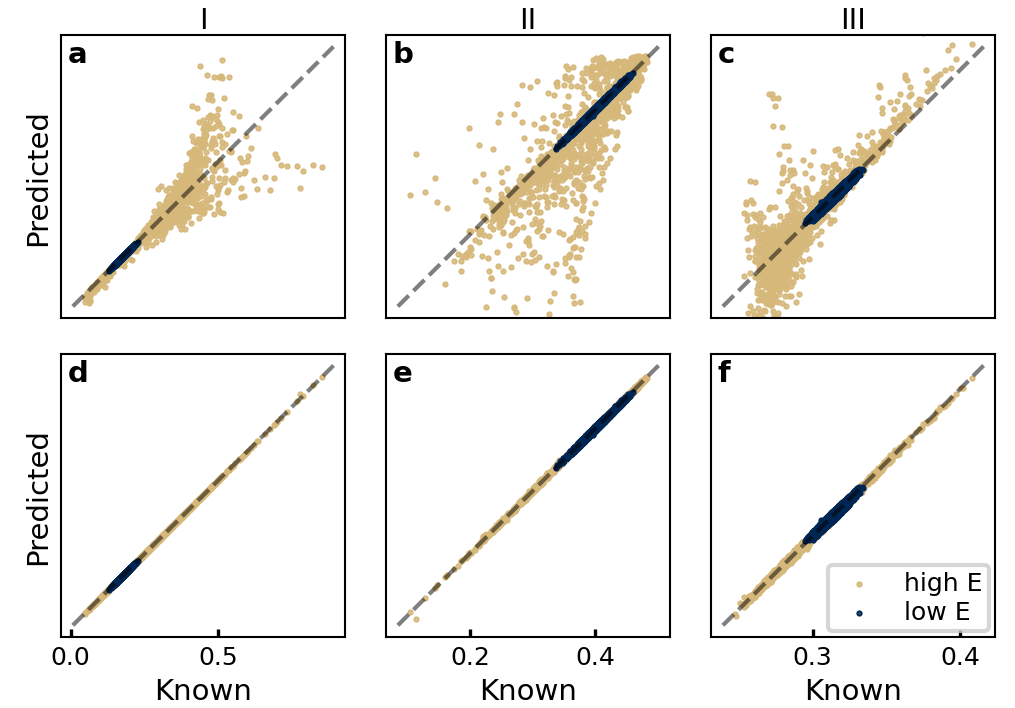}
\caption{\label{paradigm1} Spectrum ROI prediction for the two test sets presented in different colors. Model trained with low-E data (a,b,c) and model trained with high-E data (d,e,f). Complete match is represented by the gray line.}
\end{figure}

Spectrum ROI-intensity prediction via polynomial models is more accurate than any NN configuration. Given the dimensions of the argument vector, the studied polynomial models are always overdefined. For an accurate model, we observe an expected feature of bad generalization outside the training data, as shown in Figure \ref{paradigm1}. A good correspondence with the test set data is obtained by training with data of the same spread.
\par
\begin{figure}
\includegraphics[width=\linewidth]{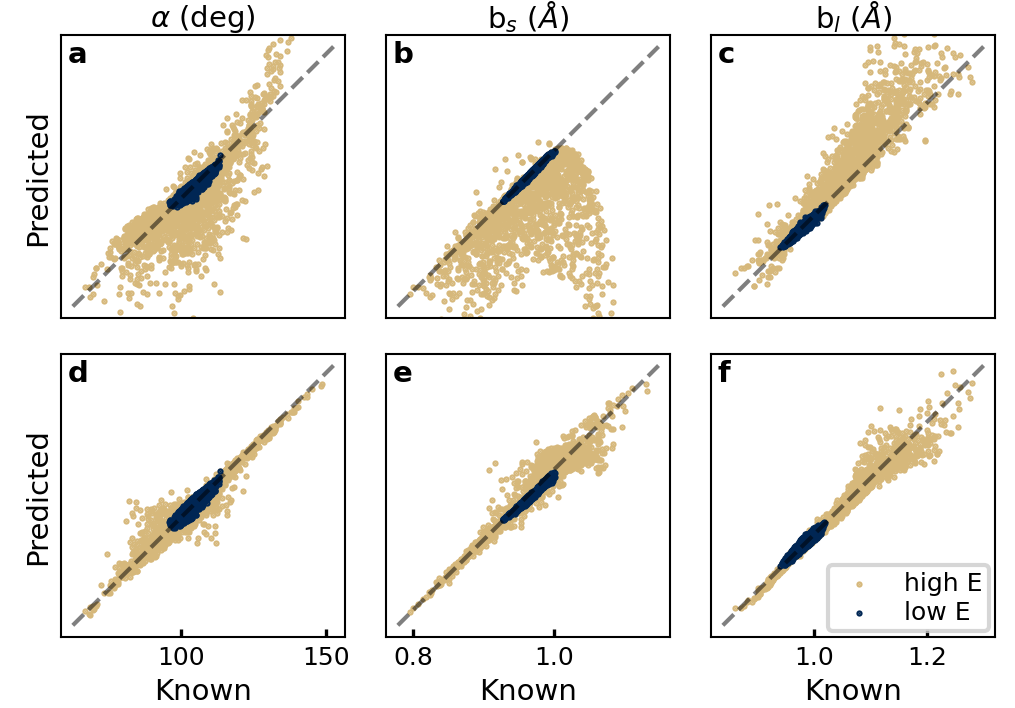}
\caption{\label{paradigm2} Structure prediction for the two test sets presented in different colors. Model trained with low-E data (a,b,c) and model trained with high-E data (d,e,f). Complete match is represented by the gray line.}
\end{figure}

For structure prediction from the 3 ROIs, NNs are more accurate (Table \ref{tab:best_comparison}) than polynomials. Even with training data covering the region of prediction, performance on the level of the corresponding spectrum ROI prediction is not observed, although the match is improved compared to training with a set of smaller coverage (Figure \ref{paradigm2}).
\par
Unlike the studied polynomial models, for the studied NN architectures the number of free parameters can exceed that of the training points. Here, among the top-performing models we found architectures of notably different complexity, especially for low-E structure prediction. For example, a NN built of 3 layers with 100 neurons in each ($\sim$20000 trainable parameters) can show the same performance as the network of 2 layers with 500 neurons ($>$ 250000 trainable parameters).
\par
\begin{figure}[ht]
\includegraphics[width=\linewidth]{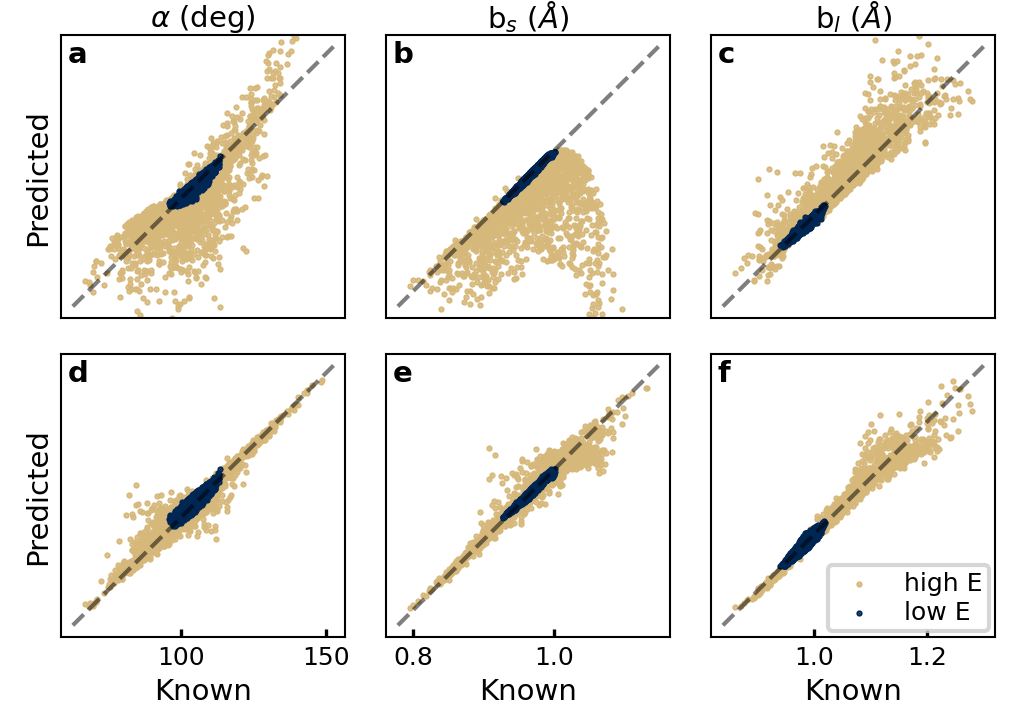}
\caption{\label{limited_dof_3x3}Free-parameter-limited structure prediction for the two test sets presented in different colors. Training with low-E data $3\times100$ layers, 20903 parameters (a,b,c). Training with high-E data $3\times100$ layers, 20903 parameters (d,e,f). Performance equal to the unlimited case can be obtained using models with limited number of parameters. Complete match is represented by the dashed line.}
\end{figure}

 
Although the presented top performers do not show drastic overfitting in the respective learning curves (see the Appendix), we investigated the NN configurations which have smaller number of parameters than the number of training samples themselves. The results of this study are presented in  Figure~\ref{limited_dof_3x3}. From the average cross-validation MSE scores (Table \ref{tab:best_comparison}), as well as from the plot, we conclude that a performance similar to the unlimited-parameter-number case can be obtained. From the learning curves (Appendix) we conclude quite similar behavior of the parameter-limited models: a few thousand training samples are needed when spectral data was coarsened to three regions of interest.
\par 
\subsection{Tight-gridded spectral data}
\begin{figure}[ht]
\includegraphics[width=\linewidth]{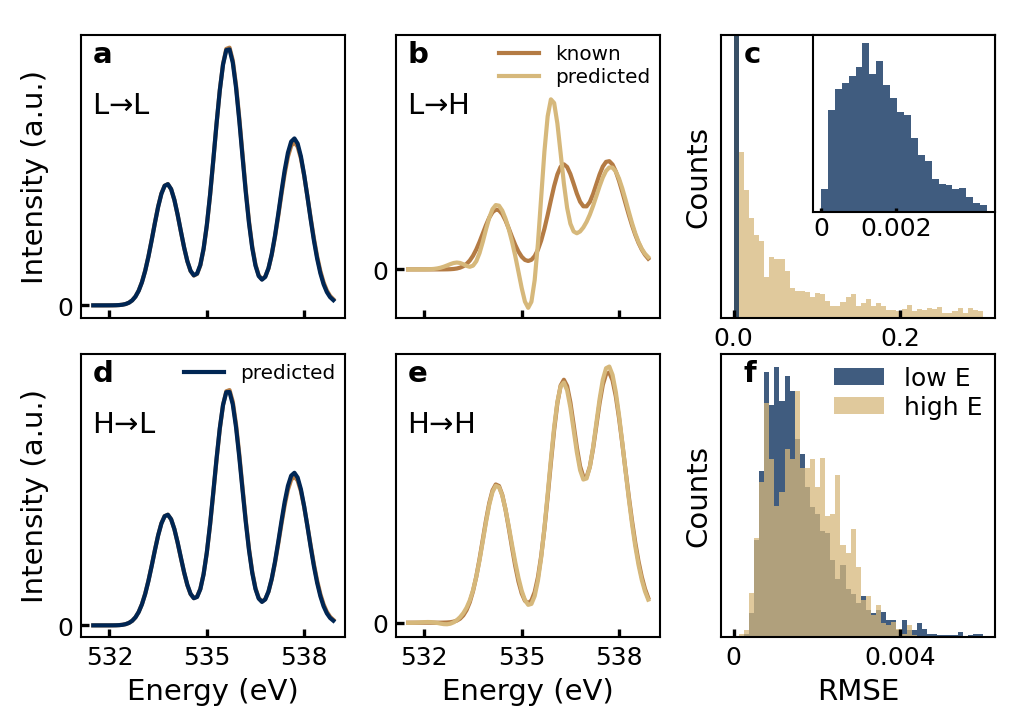}
\caption{\label{fig:si_75_par-s}Full spectrum prediction by generalized polynomial model for a sample spectrum (a, b, d, e). Labels in the top left corner of each panel indicate the prediction model, e.g. H$\rightarrow$L shows low-energy spectrum predicted by the high-energy-trained model. ~Distributions of RMSE for all predicted spectra with target value scaling (c,~f).
}
\end{figure}

Polynomial models, as implemented in this work, are independent for each target parameter, whereas NNs are not. As polynomials excelled in predicting the spectral channel intensity for the 3-ROI case, it is not tremendously surprising to find polynomials similar to those for the coarsened data to perform best, as summarized in Table~\ref{tab:best_comparison}. Figure \ref{fig:si_75_par-s} depicts sample predicted spectra using either the low-E or high-E training data sets. In addition histograms of root-mean-square errors for the test set are depicted. As in the case for the coarse-gridded data, good prediction is obtained using polynomials with adequately broad training sets. We note however, that for a tight-gridded spectrum generalized high-E spectra are somewhat better produced by NN models from the low-E training data, i.e. NN generalize slightly better. 
\par
\par
\begin{figure}[ht]
\includegraphics[width=\linewidth]{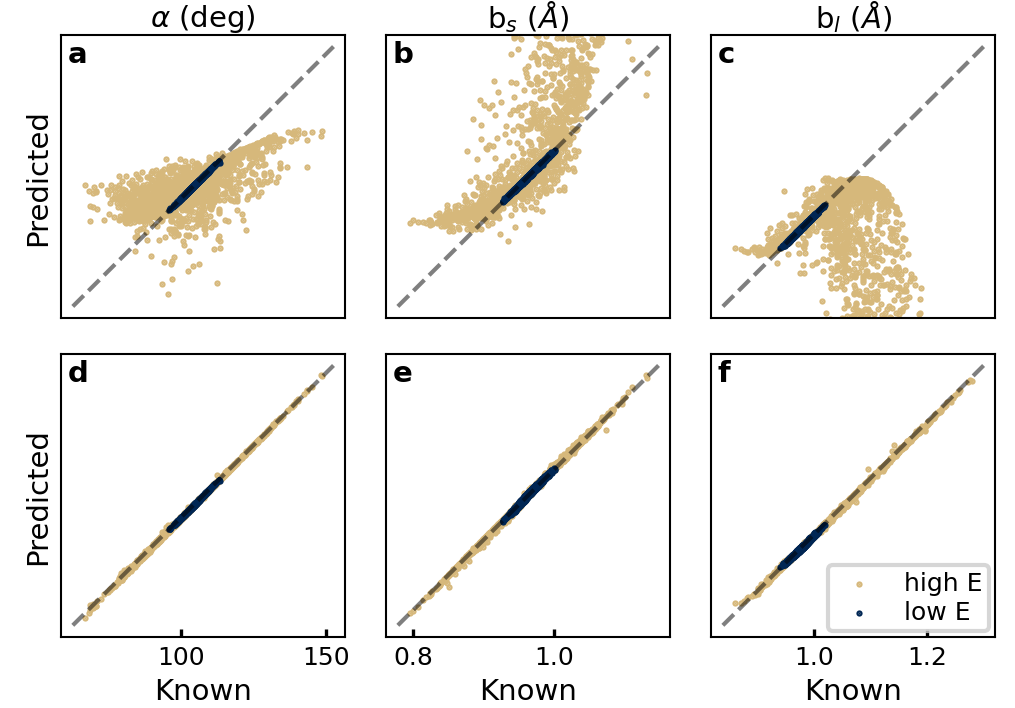}
\caption{Structure prediction for the two test sets presented in different colors for tight-gridded spectra. Model trained with low-E data (a,b,c) and model trained with high-E data (d,e,f). Complete match is represented by the gray line.}
\label{fig:si_75_s-par_unlimited}
\end{figure}
\begin{figure}[ht]
\includegraphics[width=\linewidth]{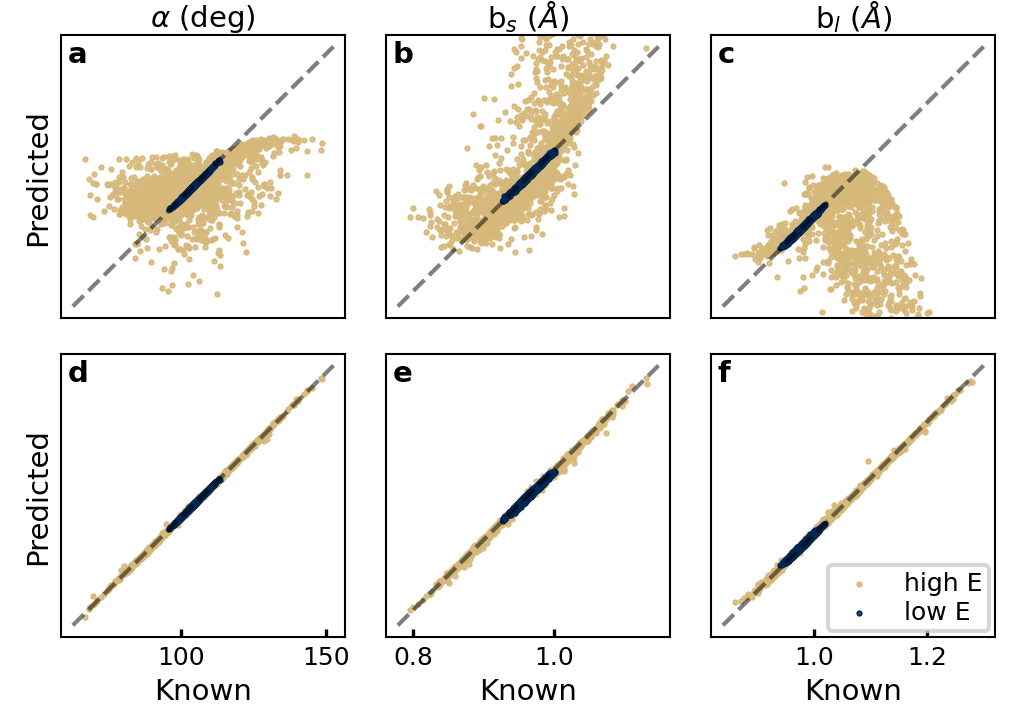}
\caption{Free-parameter-limited structure prediction for the two test sets presented in different colors for tight-gridded spectra. Model trained with low-E data (a,b,c) and model trained with high-E data (d,e,f). Complete match is represented by the gray line. Performance equal to the unlimited case can be obtained using models with limited number of parameters.}
\label{fig:si_75_s-par_limited}
\end{figure}

However, polynomials contain cross terms of the input features. Therefore the number of the free coefficients grows rapidly with their number, causing a drawback for prediction of structure. For structure prediction we studied the polynomial models up to the second order and did not limit the architecture of the neural network. In all 4 cases of structure prediction (low-E/high-E,limited/unlimited), we observed better accuracy with NN-based models (see Table~\ref{tab:best_comparison} for details). The results for structural prediction are presented in Figure \ref{fig:si_75_s-par_unlimited} and for the limited-parameter case in Figure \ref{fig:si_75_s-par_limited}. Again, sufficient training set spread plays a crucial role on the goodness of the prediction.
\par
Training of NNs to their full potential with tight-gridded spectra may require a slightly larger dataset when compared to the coarse spectra (see the learning curves in Appendix). Moreover, we find a model with limited number of parameters, that has performance similar to the optimal choice from our full grid search. Analogously to the integrated-ROI case, the training data set must cover the points of prediction. Most interestingly for structural prediction, we find extremely good performance with tight-gridded spectra, when compared to the ROI intensities.
\section{Discussion}
Owing to the Taylor expansion, polynomial models are typical approximators in physics, and sometimes coined a particular physical property (such as `heat capacity'). The drawback of polynomials is their poor extrapolation capability, especially of the higher degree models. In the models of this work, the number of degrees of freedom of polynomials explodes with increasing number of input features, but the model is not sensitive to the number of target variables. 
\par
For an alternative approach, we also used multi-layered perceptrons (MLP) \cite{deepreview2015,goodfellowbook2016} of 2 to 5 layers, which was motivated by their known ability to capture and approximate nonlinear behavior. In many learning tasks, MLPs are nowadays common, including physical sciences \cite{carleo2019,Iten2020,vladyka2020}. The classic MLP serves as a reasonably well understood test case for a neural network, and its properties have also been studied analytically for a long period. It has, for example, been shown that multilayer feedforward networks are universal approximators with saturating activation functions \cite{hornik1989,cybenko1989} and also with a wider set of activation functions \cite{leshno1993}. Even in the conceptually simple feedforward networks there are crucial free parameters such as its architecture and the chosen activation functions. The potential for using such a network for interpretation of quantum-transitions is clear: it is able to learn complex, non-linear behavior from a limited set of data.
\par
By using a test set separate from training and model selection data, independent `correct' data points are used for evaluation of prediction goodness. We observe that for good performance, the training data must cover the region for which predictions are made. Indeed, predicting the low-E values works well for all cases, as the training set is from equal or from larger cover of configuration space. When predicting the high-E case, we observe that the low-E-trained model can not generalize well.
\par
We also observe that spectrum prediction is easier than structure prediction. A potential reason for this is that while there is a function (Hamiltonian and its spectrum) from structure to spectrum, the inverse function is not guaranteed to exist. Moreover, the degree of spectrum coarsening has a significant effect on the spectrum-to-structure predictions, and thus crucial detail may be hidden from the learner by inadvertently combining separable features. As a counter argument, integration to broader ROIs captures considerable detail from the underlying system and sums over spectral details (e.g. vibrational profiles) that are not produced with the adequate level of theory. Indeed, such calculations are typical for X-ray spectra in the condensed phase. Furthermore, spectra can be simulated with arbitrarily tight grid spacing beyond any experimental meaning, which should not affect the analysis. Even though superior structure-prediction performance is obtained by tight-gridded spectra, it can be argued that this might be sensitive to the \textit{ad hoc} line shape used for convolution, which could cause biases (for discussion about the lineshape in XAS of H$_2$O, see {\it e.g.} Ref. \citenum{Leetmaa2010}). Details of the sensitivity of structural prediction by ML to the used line shape is left for future work.
\par
In the ideal case, a model trained with a limited set of samples would predict accurately beyond its training set. Good generalizability would obviously be beneficial, because sufficient coverage of relevant structures of the training set may be difficult to guarantee {\it a priori}. For this purpose, we studied the generalization performance of the structure-prediction approach. We quantify error of prediction for a data point against its known distance from the low-E mean value ${\bf P}_\mathrm{cen}$ in the units of standard deviations ($\sigma_i$) of the parameters $P_i$ in the low-E training data. For points ${\bf P}$ and ${\bf P}'$ in the structural parameter data, we defined this deviation $\chi$ as mean-absolute deviation (MAD):
\begin{eqnarray}
\chi({\bf P},{\bf P}') &=& \frac{1}{N}\sum_{i}\left|\frac{P_i-P_i'}{\sigma_{i}}\right|
\end{eqnarray}
where $\sigma_{i}$ are their standard deviations in the low-E training set ($i=\alpha,b_s,b_l$, $N=3$). This choice allows for studying the high-E set in the same units, in which the high-E data set has larger spread.
\par
\begin{figure}
\includegraphics[width=\linewidth]{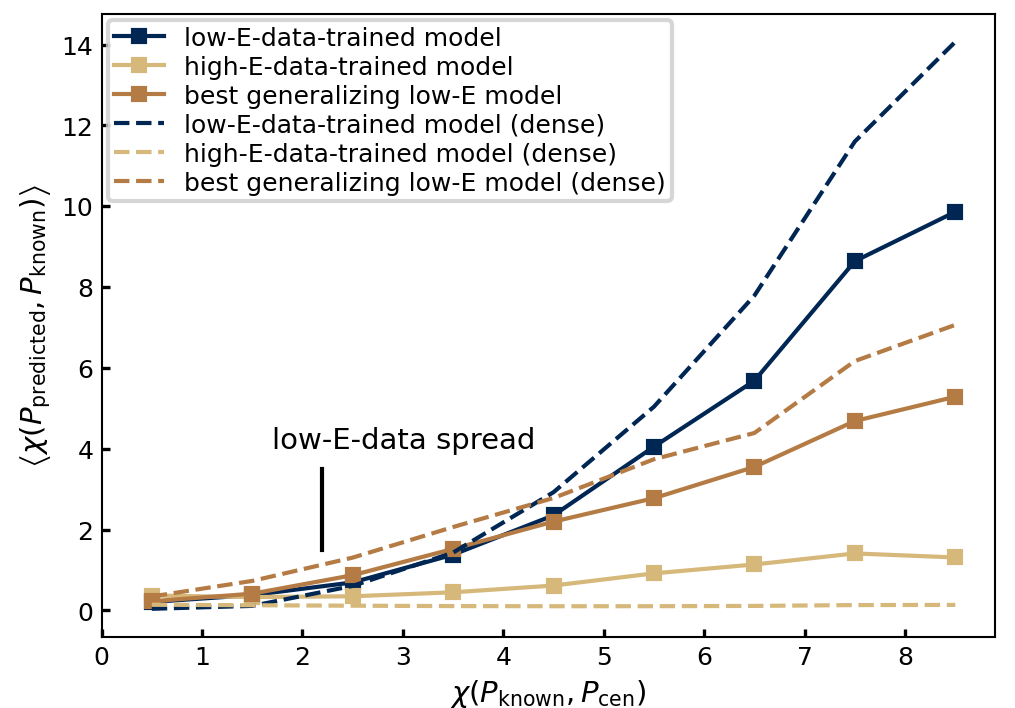}
\caption{\label{TrustRadius} Mean structure prediction error (in low-E training-set standard deviations) of the parameters as a function of deviation of the known structure from the low-E training set mean. The performance of the models for tight-gridded spectral data are marked with dashed lines.}
\end{figure}
\par
Figure \ref{TrustRadius} shows that with the chosen metrics the error for structural prediction grows with distance from low-E training set mean. As expected, use of the broader (high-E) training set results in superior performance, especially when tight-gridded data is used. The latter also holds for low-E-trained models set within their training set.
\par
The apparent problem with generalizability could, in principle, be solved by model selection. We studied this idea by selecting the low-E-trained model with the best mean squared error on high-E training data. We find the performance obtained is somewhat better than that of the low-E model, and list the best performing models in table \ref{tab:bestgen}. We observe that the best generalizing models are simpler than their counterparts used within the training set; the intuitively clear phenomenon that complicated models are best for interpolative use, whereas simpler models generalize best (i.e. least badly).
\par
\begin{table}[ht]
\caption{\label{tab:bestgen}The best-generalizing models for the two paradigms.}
\begin{tabular}{l l l l}
Prediction & Best Model\\
\hline
Spectra   & 2nd order polynomial, $\alpha=10^{3}$\\
Structure & 4 $\times$ 10 unit ReLU, $\alpha=10^{-6}$\\
Spectra (tight grid) & 1st order polynomial, $\alpha=10^{4}$\\
Structure (tight grid) & 3$\times$5 ReLU, $\alpha=10^{-4}$\\
\hline
\end{tabular}
\end{table}
\par
Rules or best practices of sampling the phase space for ML of core-level spectroscopy have not been agreed about yet, as the task is different to sampling a given statistical ensemble. In this work we opted to run the simplest simulations (NVE) to produce data of different configuration-space coverage, but see no objections to use thermostatted AIMD for the canonical ensemble. For a given pressure or volume, temperature increase may indeed provide a way to sample a large enough configuration space for a training set of sufficient coverage. However, reasonable sampling may also be hindered for other reasons, such as unaccounted quantum-probability-distribution effects, or simply because the numerical simulation might not visit allowed relevant structures, {\it e.g.} a neighborhood of a transition state. For example, using a variety of thermodynamic parameters, applying path-integral formalism and metadynamics may be needed for MD to provide enough configurational coverage to reliably interpret the spectra of the system.
\par
It might be an appealing idea to teach an ML system and then apply it for prediction of structures in the experiment. We see a large potential for error in this kind of mixed approach. First, such an idea is intrinsically assuming that spectrum evaluation (for ML learning) is accurate enough for the system not to distort the analysis. Second, ensemble averaging effects will pose a hard problem, that may potentially be underdetermined (due to number of possible structures) with any number of ROIs. Third, the experiment comes with instrumental errors, biases, and statistical noise, which could affect the outcome of the interpretation. To us it remains an open question, how surmountable these conditions are for reliable interpretation of experiments with ML trained by simulated data. As a safer route towards structural interpretation of spectral trends we propose the following method: (i) simulate over sufficient configuration space and (ii) train an ML model based on this data. Now the lightness of evaluation of the model allows for iterative algorithms to (iii) optimize the parameters of a model structural distribution for an ensemble-averaged simulated spectrum. Last, to avoid the mismatch of experiment to skew interpretation, we propose to (iv) apply changes to the simulated ensemble averaged spectrum (as seen from the corresponding experiment) and observe changes in the predicted structural distribution. This could potentially be done for example by a {SpecSwap}-{RMC} \cite{Leetmaa2010b,Zhovtobriukh2019, Zhovtobriukh2019b,Pettersson2022} approach.
\section{Conclusions}
Collaborative action of structural parameters in core-level excitation spectra of discrete transitions can be captured by the simplest machine learning applications. For XAS of the H$_2$O molecule this holds both for the prediction of a spectrum (or region of interest therein) and the prediction of a structure, as long as the training set covers the portions of phase space within which predictions are made. The latter shortcoming could not be cured by model selection based on best generalizability. For the interpretation of spectra from more complicated systems, this coverage may be hard to guarantee or prove, but variants of molecular dynamics may constitute a feasible means of covering sufficiently large yet physically meaningful portions of parameter phase space.
\par
The task of predicting spectra was simpler than predicting structures, in agreement with the fact that to this direction of prediction a function is guaranteed to exist, via solution of the Schr\"odinger equation. This finding favors statistical interpretation algorithms that root on repeated prediction of spectra rather than structure. In the test system, we even found polynomial models to outperform neural networks in the former question, and vice versa for the latter one. Last, we found structure-predicting machine learning models operating with tight-gridded spectra to outperform ones with coarse grid with the same data. This may invite for over-interpretation, as it is possible to present simulation results on an arbitrarily dense grids, beyond experimental resolution or accuracy of the calculation.

\section{Acknowledgements}
JN and AV acknowledge Academy of Finland for funding via project 331234.

\bibliography{references}
\FloatBarrier
\newpage
\newpage

\appendix*
\section{learning curves}

\begin{figure}[ht]
\includegraphics[width=0.95\linewidth]{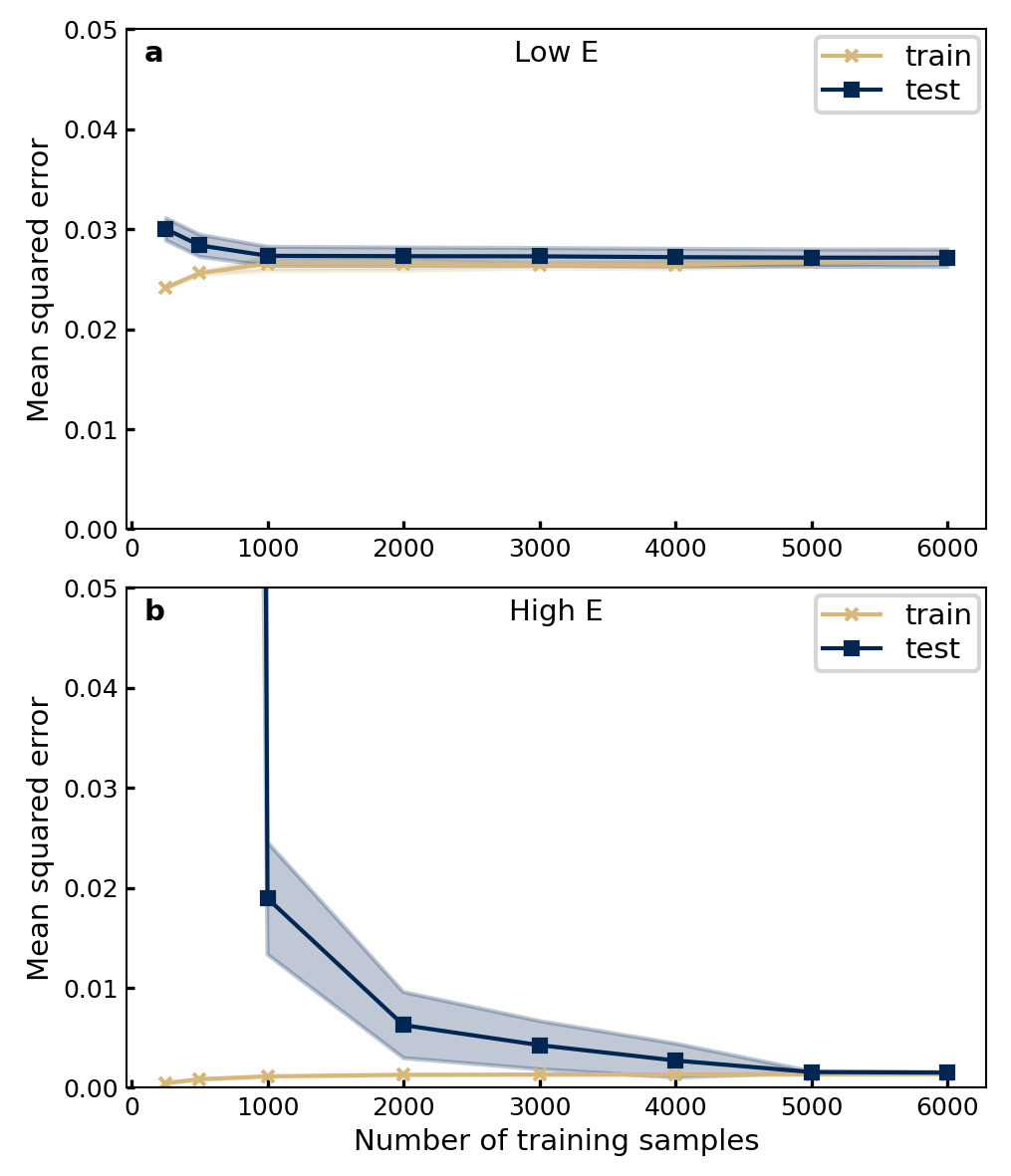}
\caption{Structure to spectrum learning curves for the parameter-number-unlimited ROI-integrated case.}
\end{figure}

\begin{figure}[ht]
\includegraphics[width=0.95\linewidth]{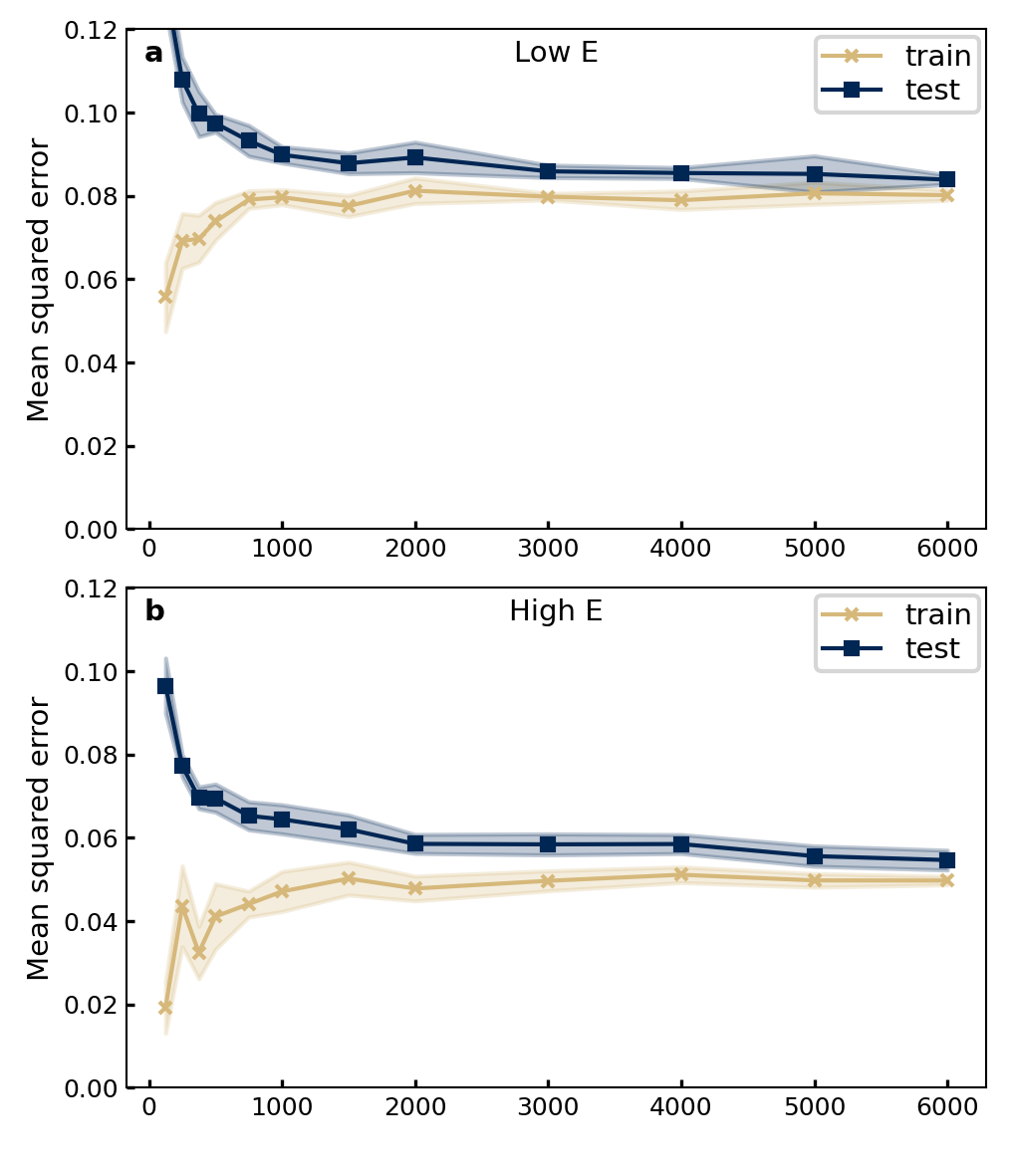}
\caption{Spectrum to structure learning curves for the parameter-number-unlimited ROI-integrated case.}
\end{figure}

\begin{figure}[ht]
\includegraphics[width=0.95\linewidth]{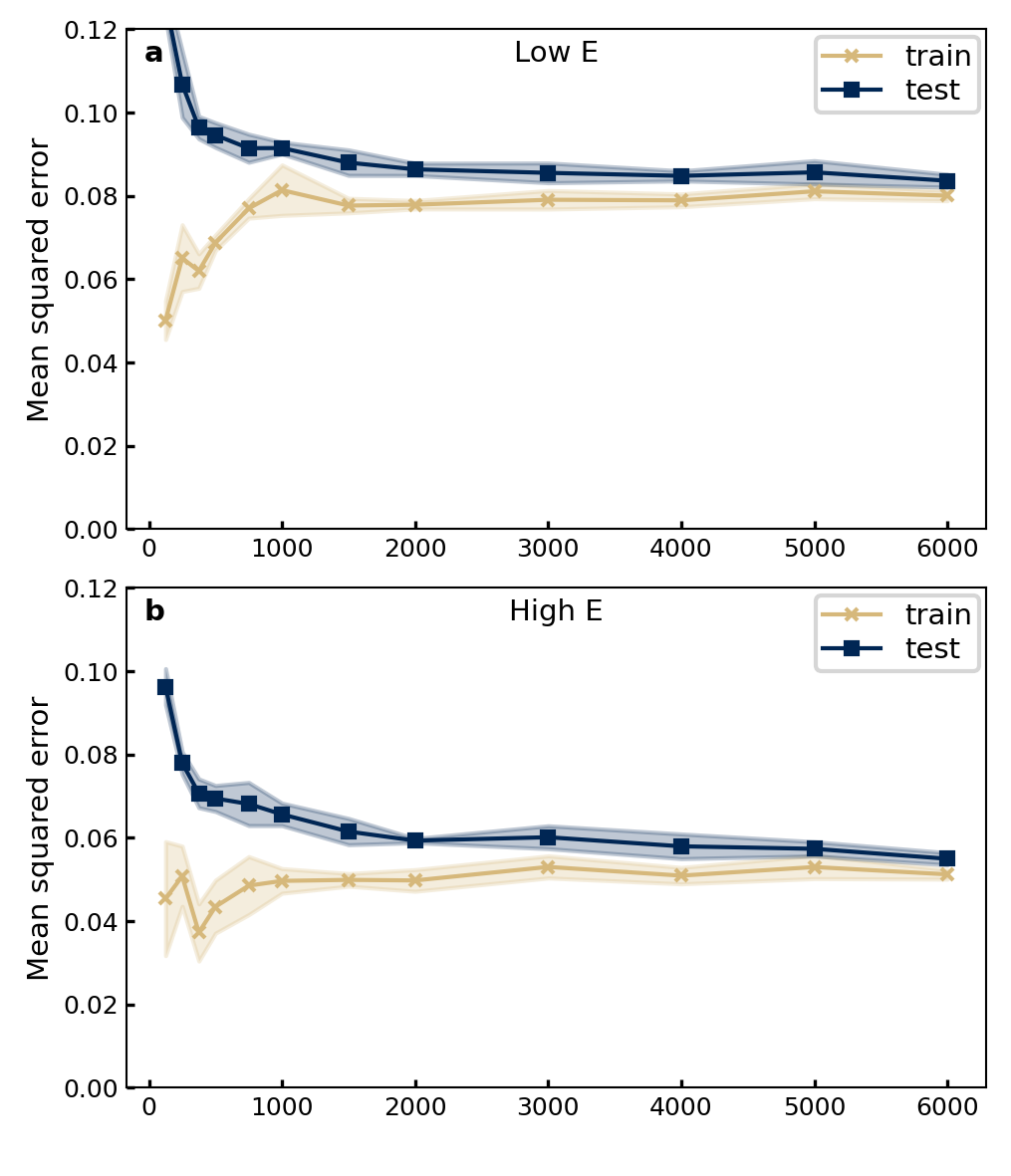}
\caption{Spectrum to structure learning curves for the parameter-number-limited ROI-integrated case.}
\end{figure}

\begin{figure}[ht]
\includegraphics[width=0.95\linewidth]{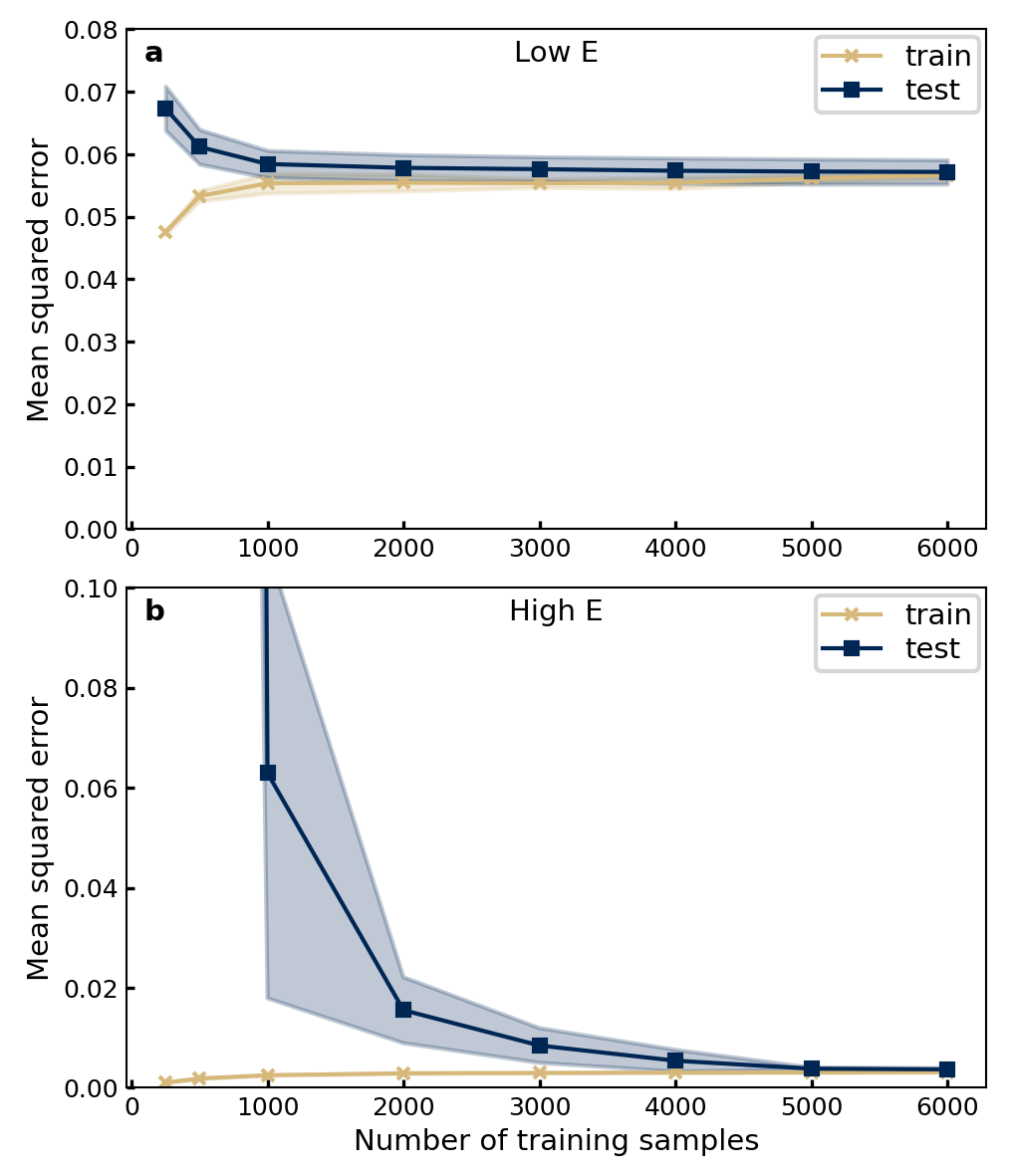}
\caption{Structure to spectrum learning curves for the tight-gridded case, polynomial models.}
\label{fig:learning:75:spectrum}
\end{figure}

\begin{figure}[ht]
\includegraphics[width=0.95\linewidth]{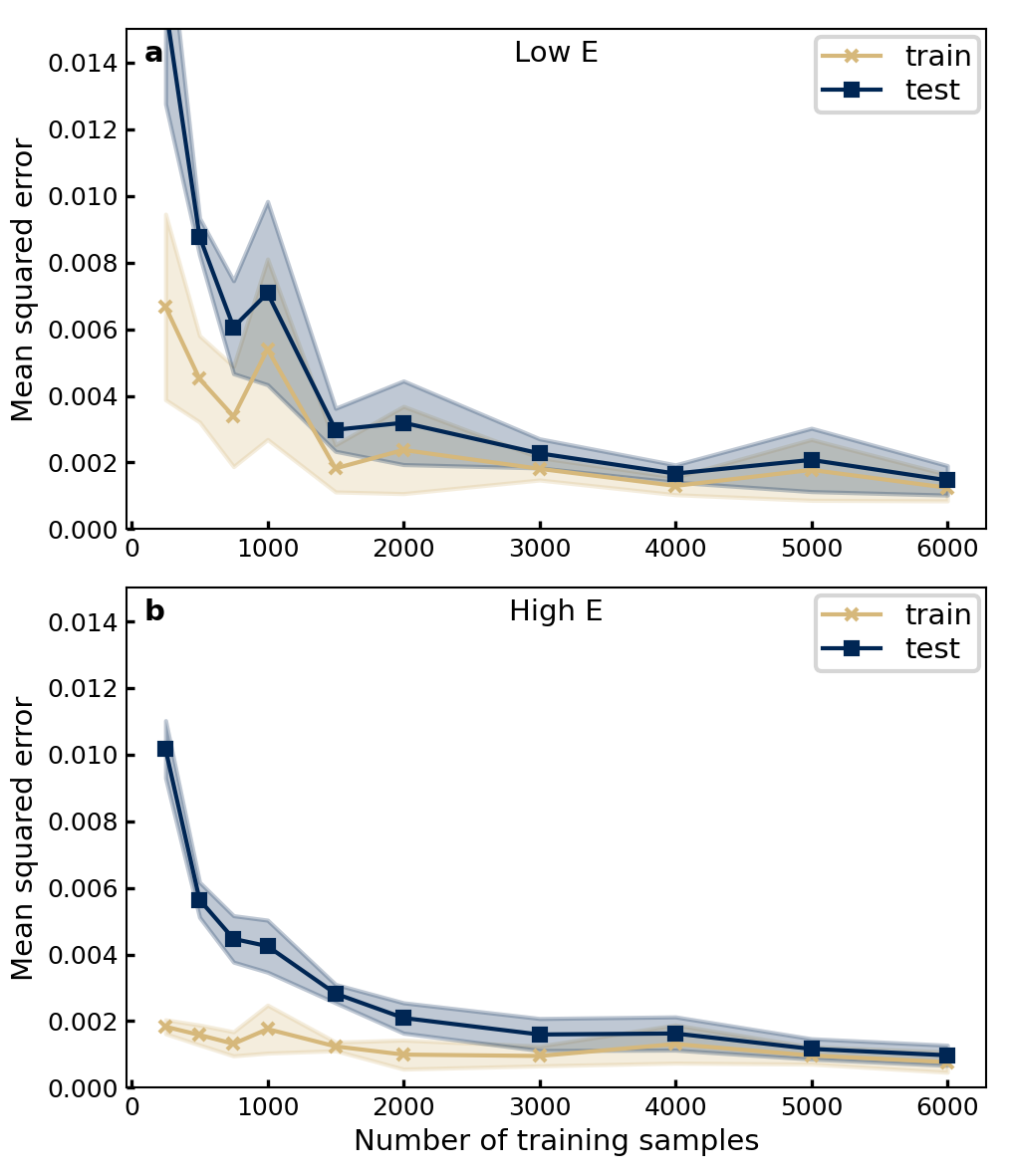}
\caption{Spectrum to structure learning curves for the parameter-number-unlimited tight-gridded case.}
\label{fig:learning:75:unlim}
\end{figure}

\begin{figure}[ht]
\includegraphics[width=0.95\linewidth]{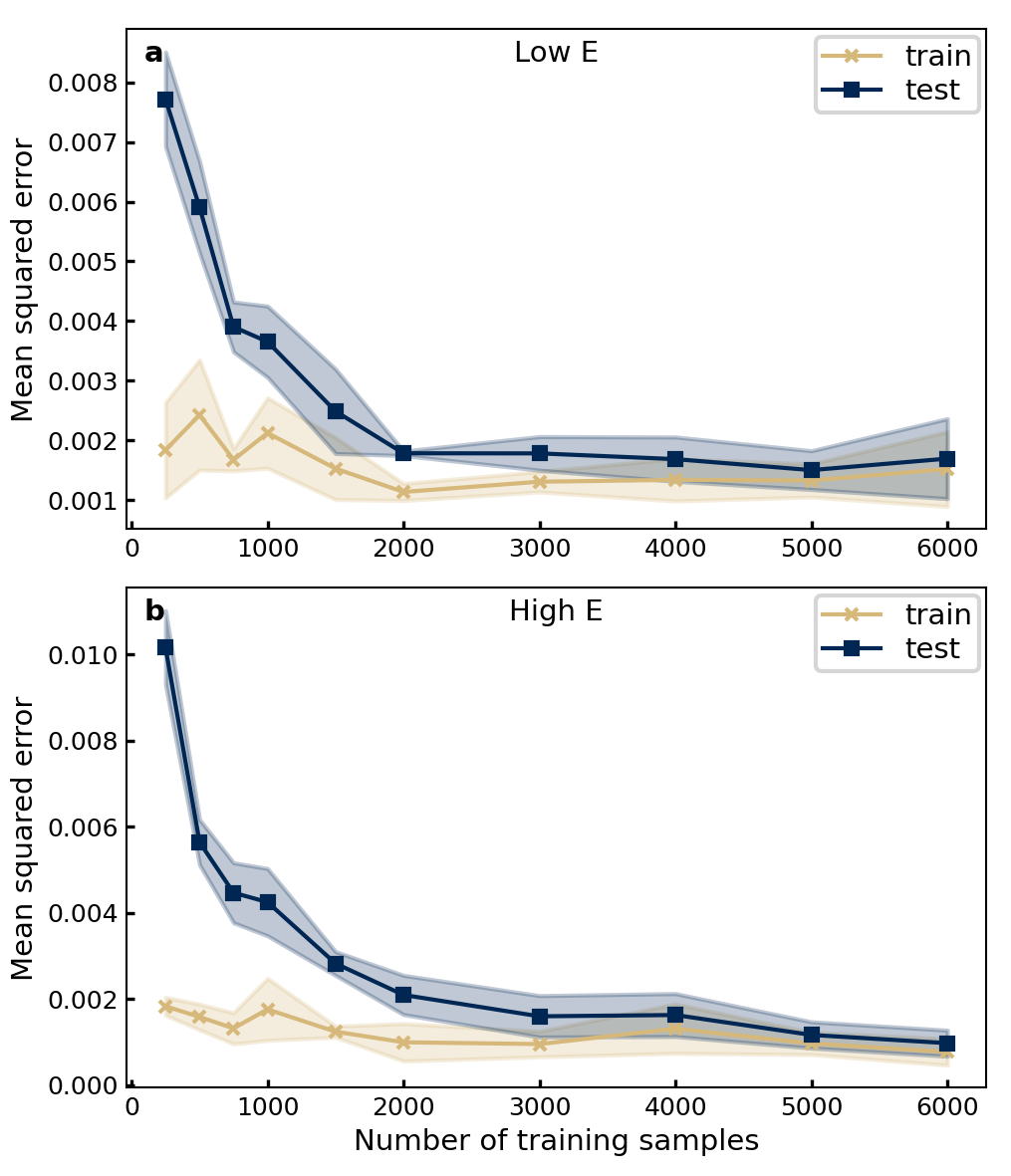}
\caption{Spectrum to structure learning curves for the parameter-number-limited tight-gridded case.}
\label{fig:learning:75:lim}
\end{figure}

\end{document}